\documentclass[aps,pre,amsmath,amssymb,reprint,onecolumn]{revtex4-1}
\usepackage[normalem]{ulem}
\usepackage[shortlabels]{enumitem}
\usepackage{CJK}
\usepackage{amssymb,euscript,latexsym,amsmath}
\usepackage{graphicx}
\usepackage[dvips]{epsfig}
\usepackage{color}
\usepackage{epstopdf}
\usepackage{setspace}
\usepackage{subfigure}
\usepackage{chngpage}
\usepackage{hyperref}

\usepackage[T1]{fontenc}
\usepackage[utf8]{inputenc}
\usepackage[font=small,labelfont=bf,tableposition=top]{caption}
\usepackage{booktabs,multirow}

\DeclareGraphicsExtensions{.eps,.ps,.pdf}

\renewcommand{\v}[1]{\ensuremath{\mathbf{#1}}} 

\begin{document}

\title{Mucociliary Transport in Healthy and Diseased Environments}

\author{Hanliang Guo}
\author {Eva Kanso}
\email{kanso@usc.edu}
\affiliation{Department of Aerospace and Mechanical Engineering,  \\ University of Southern California, Los Angeles, California 90089, USA}

\date{June 6, 2016}

\begin{abstract}
Mucociliary clearance in the lung is the primary defense mechanism that protects the airways from inhaled toxicants and infectious agents. The system consists of a viscoelastic mucus layer on top of a nearly-viscous periciliary layer surrounding the motile cilia. In healthy environments, the thickness of the periciliary layer is comparable to the cilia length.  Perturbations to this system, whether due to a genetic disorder or acquired causes, are directly linked to infection and disease. For example, depletion of the periciliary layer is typically observed in diseases such as chronic obstructive pulmonary disease and cystic fibrosis. Clinical evidence connects the periciliary layer depletion to reduced rates of mucus clearance.  In this work, we develop a novel computational model to study mucociliary transport in a microfluidic channel. We systematically vary the viscoelastic properties and thickness of the mucus layer to emulate healthy and diseased environments. We assess cilia performance in terms of three metrics:  flow transport, internal power expended by the cilia, and transport efficiency. We find that, compared to a control case with no mucus, a healthy mucus layer enhances cilia performance in all three metrics. That is to say, a healthy mucus layer not only improves flow transport, resulting in better clearance of harmful substances, but it does so at an energetic advantage to the cilia. Stiffer mucus  enhances further the transport efficiency. In contrast, in diseased environments where the periciliary layer is depleted, mucus hinders transport and stiffer mucus leads to a substantial decrease in transport efficiency.  This decrease in transport is accompanied by an increase in the internal power needed to complete the cilia beating cycle. Cilia failure would occur when the required power is higher than that afforded by the cilia internal machinery. 
This work provides a quantitative framework for assessing cilia performance in flow transport in healthy and diseased environments, that can be equally applied to experimental data on cilia-driven flows. It can thus provide a tool for the diagnostic of diseases related to mucociliary transport. 

%
\end{abstract}

%
\maketitle

\singlespacing

\section{Introduction}
Cilia are microscopic hairlike organelles present on many cells in the mammalian body, either in large groups on a single cell or as solitary structures~\cite{Christensen2007}. 
Motile groups of cilia, the focus of this work, are found on the epithelial cells of the trachea~\cite{fulford1986, ocallaghan_respiratory_1999,randell_effective_2006}, ependymal cells in the brain~\cite{DelBigio1995,Mirzadeh2010} and  cells lining the oviduct and epididymis of the reproductive tracts~\cite{Lyons2006}. They normally beat in an asymmetric pattern resulting in fluid movement and transport of particles and cells~\cite{wong_nature_1993}.  Dysfunction in the ciliary machinery results in decreased transport rates and can lead to disease and impaired organ function\textcolor{black}{~\cite{davenport_incredible_2005,brooks2014multiciliated}}.

\begin{figure}[t]
\centerline{\includegraphics{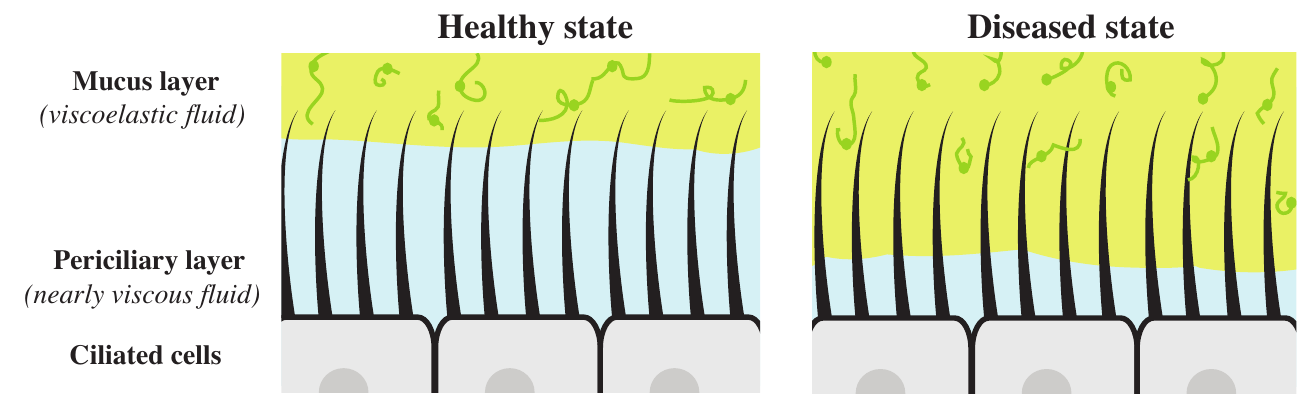}}
\caption{{Healthy and diseased ciliary systems.} In a healthy state,  cilia lie mostly in a nearly-viscous periciliary layer, with only the tips penetrating into the viscoelastic mucus layer. In a diseased state, the mucus is denser and thicker, submerging the cilia in a mostly viscoelastic environment. }
\label{fig:intromodel}
\end{figure}

In the mammalian lung for example, mucus clearance is the primary defense mechanism that protects the airways from inhaled toxicants and infectious agents~\cite{cone2009, boucher2007}. 
Failure of mucus clearance is linked to  human
lung diseases such as chronic obstructive
pulmonary disease (COPD)~\cite{donno2000, rogers2004, rogers2005, hogg2004, comer2012, seys2015} and cystic fibrosis
(CF)~\cite{boucher2007, livraghi2007, wielputz2013}. 
The mucociliary clearance system 
consists of two components: a viscoelastic mucus
layer that traps inhaled particles and gets transported
out of the lung by cilia-generated forces, and
 a low-viscosity periciliary layer that facilitates the beating of cilia; see Fig.~\ref{fig:intromodel}(a). 
 Great advances have been made in understanding the mechanics of ciliary transport and cilia-generated flows; see, for example,~\cite{chopra_measurement_1977,smith_modelling_2008, smith_mathematical_2009,li_methods_2012,ding2014}.
However, there is a shortage of quantitative models that
predict the degree of failure in mucus transport under perturbed and diseased conditions; Fig.~\ref{fig:intromodel}(b). 
Predictive airway clearance
models 
would  improve the understanding of cilia-related lung diseases and the development of treatment therapies~\cite{boucher2004, wanner_mucociliary_1996, button2012}.


Early mathematical models of mucociliary transport date back to the work of Barton and Raynor (1967). The authors considered the cilium to be a rigid rod that is shorter during the recovery stroke than during the effective stroke, and approximated its effect on the surrounding fluid using  ``resistance" coefficients that allowed them to obtain somewhat realistic flow rates~\cite{barton1967}. Numerous studies  were conducted thereafter, shedding more light on the mechanics of mucociliary transport; see, for example,~\cite{fulford1986, smith_modelling_2008, lee2011, jayathilake2012, montenegro2013, jayathilake2015, li2016, chatelin2016} and references therein. Most of these works consider single or two-layer Newtonian fluids, focusing on the effects of viscosity on fluid transport.
Viscoelastic properties of the mucus were first taken into account by Ross (1971). Ross considered a Maxwell fluid propelled by a continuous ``wavy wall" and analytically solved for the fluid transport rates~\cite{ross1971}. More recently, Smith {\em et al.} (2007) proposed a traction layer model consisting of three layers: a periciliary layer of Newtonian fluid, a mucus layer of Maxwell fluid, and a thin traction layer between the periciliary and mucus layers accounting for the interaction between the cilia and the mucus. The effect of cilia beating is modeled as a time-dependent force acting on the traction layer~\cite{smith2007viscoelastic}. The traction-layer model predicts physiologically-reasonable values of mucus transport and provides great insight into the temporal and spatial details of mucociliary flows. However, it  is limited in that it cannot take into account the details of specific cilia beating patterns. It is also limited by the Maxwell fluid assumption, which is a linear model of viscoelasticity.
 \textcolor{black}{It is worth noting here that the periciliary layer was generally considered as a Newtonian fluid layer, without elasticity or other complex properties, but recent investigations suggest that this view may be too simplistic~\cite{tarran2001,button2012}.}

In this paper, we present a mucociliary transport model consisting of two fluid layers: a mucus layer of highly viscoelastic fluid on top of a periciliary layer of nearly viscous fluid, as in the case of biological ciliary systems~\cite{boucher2004}. We account for the viscoelastic effects using the Oldroyd-B fluid model~\cite{larson1999}. The Oldroyd-B model, despite its simplicity, captures the main  mechanical features of  viscoelastic fluids under shear conditions, that is to say, under conditions that are reminiscent to  the shearing motions produced by beating cilia.  We consider a cilia beating pattern constructed from the rabbit tracheal cilia~\cite{fulford1986}.  
We  formulate a system of equations governing fluid-cilia interactions and solve it numerically using the immersed boundary method. The immersed boundary method was first proposed by Peskin (1972) to study flow around heart valves~\cite{peskin1972, peskin2002}. Since then, it was further developed and applied successfully to various fluid-structure interaction problems including problems involving non-Newtonian fluids~\cite{dillon2007, teran2010, chrispell2011shape, thomases2014}. Here, we present a systematic study of the effects of the  parameters of the mucus layer on mucociliary transport. In particular, we vary the mucus properties and thickness to emulate a range of healthy and diseased conditions. Our findings suggest that the mucus parameters greatly affect mucociliary transport. We conclude by discussing the significance of these results in relation to mucociliary transport in healthy and diseased conditions, as well as the design  of microfluidic transport mechanisms for biological and artificial cilia.

\section{Model and Method}
\subsection{Problem Formulation}
\begin{figure}[t]
	\centerline{\includegraphics{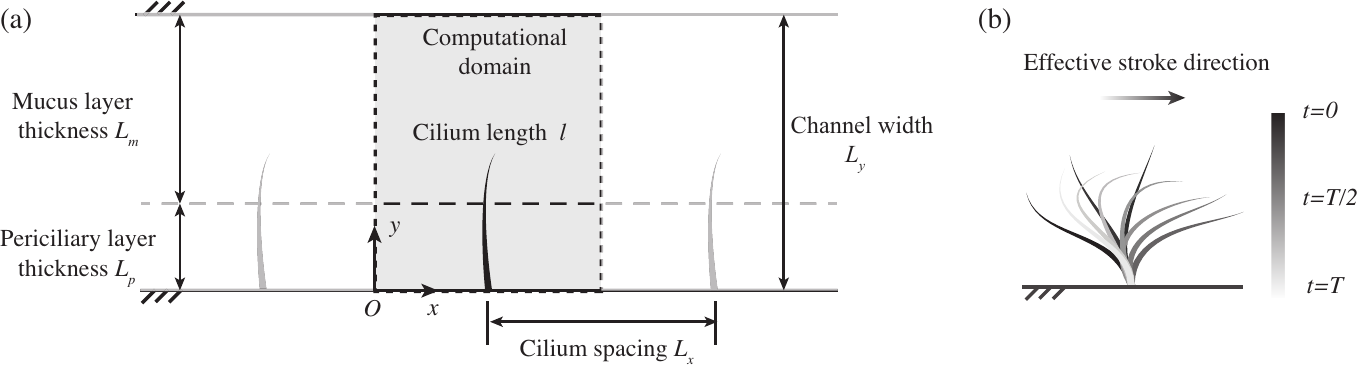}} 
	\bigskip
	\caption{{{Model schematics.} ({a}) an infinite array of motile cilia beating synchronously in a channel. The channel is filled with a viscoelastic fluid of thickness $L_m$ atop  a nearly viscous fluid of  thickness $L_p$. ({b}) Kinematics of rabbit tracheal cilia, based on~\cite{fulford1986} and scaled to preserve the length of the cilium. The effective stroke is shown in dark grey and the recovery stroke in light grey.}}
	\label{fig:compmodel} 
\end{figure}

Consider an infinite array of cilia beating in synchrony in a narrow channel of width $L_y$.  The cilia have length $l$ and are uniformly distributed on one side of the channel at a separation distance $L_x$ as shown in Fig.~\ref{fig:compmodel}(a). The channel is filled with two fluid layers: a viscoelastic mucus layer of thickness $L_{m}$ on top of a nearly-viscous periciliary layer of thickness $L_{p}$, with $L_p + L_m = L_y$. 

The cilia beating kinematics are based on  experimental data of the rabbit tracheal cilia~\cite{fulford1986} and depicted in Fig.~\ref{fig:compmodel}(b).  Mathematically, the beating kinematics can be described in a Cartesian frame $(x,y)$ attached at the based of the cilium using the vector representation $\boldsymbol{\xi}_c(s,t)$, where $s$ is the arclength along the cilium's centreline from its base $(0 < s < l)$ and $t$ is time $(0 < t < T)$. The $(x_c,y_c)$ components of $\boldsymbol{\xi}_c(s,t)$ are given by a Fourier series expansion in {$t$} and Taylor series in {$s$} with coefficients chosen to match the experimental data. 
It should be noted that the cilium length is not conserved by the coefficients reported in~\cite{fulford1986}. Here, we rescaled the coefficient to ensure the total length of the cilium is constant at all time.

The mucus and periciliary layers are described using the Oldroyd-B  model  for polymeric fluids, that is to say, for fluids consisting of a viscous fluid solvent and a polymeric elastic solute \cite{larson1999}. 
The total deviatoric stress $\boldsymbol\sigma = \boldsymbol\sigma_f + \boldsymbol\sigma_e$ of the Oldroyd-B fluid  consists of contributions from the Newtonian fluid solvent $\boldsymbol\sigma_f$ and the polymeric elastic solute $\boldsymbol\sigma_e$.
The constitutive stress-strain relations  are given by 
\begin{equation}
\boldsymbol\sigma_f = 2\mu_f \boldsymbol{D}(\boldsymbol{u}),\qquad \boldsymbol\sigma_e +r\boldsymbol\sigma_e^\triangledown - 2\mu_e\boldsymbol{D}(\boldsymbol{u})\textcolor{black}{-}\epsilon \nabla^2\boldsymbol\sigma_e = \boldsymbol{0},
\label{eq:constitutive}
\end{equation}
where $\boldsymbol{D}(\boldsymbol{u}) = \frac{1}{2}[\nabla\boldsymbol{u}+(\nabla\boldsymbol{u})^T]$ is the strain rate tensor, $r$ is the relaxation time of the viscoelastic fluid, $(\cdot)^\triangledown \equiv \frac{\partial}{\partial t}(\cdot) + \boldsymbol{u\cdot}\nabla(\cdot) - [\nabla\boldsymbol{u}(\cdot) + (\cdot)(\nabla\boldsymbol{u})^T ]$ denotes the upper convected derivative, $\mu_f$ and $\mu_e$ are the viscosities of the fluid solvent and elastic solute respectively. The relaxation time of the viscoelastic material describes {the time required for the elastic polymers in the fluid to return to equilibrium after the stress is released \cite{larson1999}. Informally, it could also be understood as ``the duration for which the material remembers the effect of an applied force.''}  Larger $r$ means that applied forces will remain effective for a longer time after unloading.
It should be noted that the diffusion term $\epsilon \nabla^2\boldsymbol\sigma_e$ is not inherent to the constitutive relation of the Oldroyd-B model. It has been added as a regularization term with $\epsilon \ll 1$ because, in its absence, the stress tensors of the Oldroyd-B model have the potential to lose smoothness in the long-time limit \cite{thomases2011, thomases2014}.

The equations of motion of the Oldroyd-B fluid are obtained by substituting the regularized constitutive relations into the balance of linear momentum. To this end, one gets the modified Navier-Stokes equation:
\begin{equation}
\begin{split}
\label{eq:NSmod}
	\rho(\frac{\partial\boldsymbol{u}}{\partial t} + \boldsymbol{u\cdot}\nabla\boldsymbol{u})+\nabla p -\mu_f\nabla^2\boldsymbol{u} - \nabla\boldsymbol{\cdot\sigma_e} -\boldsymbol{f} &= \boldsymbol{0}. 
\end{split}
\end{equation}
Here, $\rho$ is the fluid density, $p$ is the pressure field and $\boldsymbol{f}$ is the body force density acting on the fluid.  Eq.~\eqref{eq:NSmod} is to be solved 
in the fluid channel in conjunction with the incompressibility equation $\nabla \cdot \boldsymbol{u}= 0$ and the constitutive equation for 
$\boldsymbol{\sigma}_e$ from~\eqref{eq:constitutive}.
The solution to this coupled system of equations should satisfy the no-slip boundary conditions at the channel walls and along the individual cilia
\begin{equation}	\label{eq:bc}
\begin{split}
\boldsymbol{u}&= \left\{ \begin{array}{c l}
	\dfrac{\partial\boldsymbol{\xi}_c}{\partial t} &\text{at the cilia}, \\[2ex]
	\boldsymbol{0} \  &\text{on the channel walls: } y=0 \text{ and } y = L_y.
	\end{array} \right.  
\end{split}
\end{equation}

To emulate the effect of the mucus layer on top of the periciliary layer, we consider two viscoelastic fluids of different material properties.  The mucus layer consists of a  viscoelastic fluid of thickness $L_m$  and \textcolor{black}{longer} relaxation time $r_m$. The mucus layer lies  on top of a nearly-viscous periciliary fluid of thickness $L_p$ and \textcolor{black}{shorter} relaxation time $r_p$. 
Previous experimental and numerical works have shown that the interface between the mucus and periciliary layers exhibits no significant deformations under the effect of cilia beating \cite{sanderson1981,smith_modelling_2008,dillon2007}. We therefore assume no deformation at the interface. This assumption means that only the solvent could be exchanged between the periciliary and mucus layers but not the polymer molecules which dictate the relaxation time of the viscoelastic fluids~\cite{randell_effective_2006,button2012}. 
Mathematically, the fluid velocities are continuous at the interface and the extra shear stress $\boldsymbol{\sigma}_e$ is zero, $\boldsymbol\sigma_{e}|_{{y=L_p}} = 0.$

We use the cilium length $l$ to scale length, the beating cycle $T$ to scale time, the total viscosity $\mu=\mu_f + \mu_e$ to scale viscosity, and the ratio between the total viscosity of the fluid to the beating cycle $\mu/{T}$ to scale pressure. To this end, we get three non-dimensional parameters: (i) the Reynolds number $\text{Re} ={\rho l^2}/{\mu T}$  measures the ratio of inertial to viscous effects and, therefore, is negligible in drag-dominant flows; (ii) the Deborah number $\text{De} = {r}/T$  measures the elastic properties of the viscoelastic fluid and takes two distinct values: $\text{De}_m$ in the mucus layer and $\text{De}_p$ in the periciliary layer; and (iii) the ratio of elastic to total viscosity $\alpha = {\mu_e}/\mu$ is taken to be the same in the mucus and periciliary layers. 
In non-dimensional form, Eq.~\eqref{eq:NSmod} and the incompressibility condition are written as
\begin{equation}	\label{eq:govern}
\begin{split}
\text{Re}(\frac{\partial\boldsymbol{u}}{\partial t} + \boldsymbol{u\cdot}\nabla\boldsymbol{u})+\nabla p -(1-\alpha)\nabla^2\boldsymbol{u} - \nabla\boldsymbol{\cdot\sigma_e} -\boldsymbol{f} &= \boldsymbol{0}, 
\\
 \nabla\boldsymbol{\cdot u} &= 0. 
\end{split}
\end{equation}
These equations are coupled to  the constitutive relations for $\boldsymbol{\sigma}_e$ from Eq.~\eqref{eq:constitutive}, which in non-dimensional form are given by
\begin{equation}	\label{eq:govern2}
\begin{split}
\text {De}_p \ \boldsymbol\sigma_e^\triangledown + \boldsymbol\sigma_e - 2\alpha\boldsymbol{D}(\boldsymbol{u}) - \epsilon \nabla^2\boldsymbol\sigma_e &= \boldsymbol{0}, \qquad \text{for } 0\le y<L_p, \\[1ex]
\text {De}_m \boldsymbol\sigma_e^\triangledown + \boldsymbol\sigma_e - 2\alpha\boldsymbol{D}(\boldsymbol{u}) - \epsilon \nabla^2\boldsymbol\sigma_e &= \boldsymbol{0}, \qquad
\text{for }  L_p\le y<L. 
\end{split}
\end{equation}
The set of Eqs.~\eqref{eq:govern} and~\eqref{eq:govern2}  reduce to the typical incompressible Navier-Stokes equations when $\text{De}_{p} =\text{De}_{m}  = 0$, $\alpha = 0$, and $\epsilon = 0$.


\subsection{Numerical Method}


For viscous fluids ($\boldsymbol{\sigma}_e\equiv\boldsymbol{0}$) and fixed boundaries,  the incompressible Navier-Stokes equations~\eqref{eq:govern} can be discretized using a standard fluid solver, such as the classical fractional step method where pressure $p$ is treated as a Lagrange multiplier to ensure the fluid velocities satisfy the incompressibility condition \cite{chorin1968,chang2002,fletcher2012}.
In the case of moving boundaries, a standard numerical method for solving Eq.~\eqref{eq:govern} is the immersed boundary method (IBM) \cite{peskin1972,peskin2002}. IBM uses a standard Eulerian (fixed) mesh to solve for the fluid velocity field and a Lagrangian (moving) mesh to account for the moving boundary, here, the cilium. To communicate between these two meshes, a regularized Dirac delta function $\delta$ is used to project the boundary forces $\boldsymbol{F}$ onto the fluid domain,
\begin{equation}	\label{eq:interp1}
\begin{split}
\boldsymbol{f}(\boldsymbol{x},t) = \int_\mathcal{C} \boldsymbol{F}(\boldsymbol{\xi},t)\delta(\boldsymbol{x}-\boldsymbol{\xi}(s,t))\mathrm{d}s,
\end{split}
\end{equation}
and to project the fluid velocities $\boldsymbol{u}$ onto the moving boundary,
\begin{equation}	\label{eq:interp2}
\begin{split}
\frac{\partial \boldsymbol{\xi}(s,t)}{\partial t} = \int_\mathcal{F} \boldsymbol{u}(\boldsymbol{x},t)\delta(\boldsymbol{x}-\boldsymbol{\xi}(s,t))\mathrm{d}\boldsymbol{x}.
\end{split}
\end{equation}
Here,  $\mathcal{C}$  denotes  the Lagrangian cilia boundary and $\mathcal{F}$ the Eulerian fluid domain. It should be emphasized that 
$\boldsymbol{\xi}$ and $\boldsymbol{F}$ are the Lagrangian mesh markers and force density along $\mathcal{C}$. 
When the motion of the boundary is unconstrained, $\boldsymbol{\xi}$ and $\boldsymbol{F}$ are typically related via an elastic energy function $E$ such that $\boldsymbol{F}(\boldsymbol{\xi},t) = {\partial E}/{\partial \boldsymbol\xi}$. 
In this work,  the cilium $\mathcal{C}$ is constrained to follow prescribed beating kinematics $\boldsymbol{\xi}_c(t)$. 
A typical technique to link $\boldsymbol{\xi}(t)$ to $\boldsymbol{\xi}_c(t)$ is to impose
$\boldsymbol{F}(\boldsymbol{\xi},t) = K[\boldsymbol{\xi}_c(t)-\boldsymbol{\xi}(t)]$ and use large values of the stiffness parameter $K$ 
to guarantee that the motion of the boundary $\boldsymbol{\xi}$ is close to the prescribed motion $\boldsymbol{\xi}_c$.
However, this technique renders the equations of motion stiff and therefore prohibits the use of large time steps.

To circumvent this difficulty, we use the  one-way coupled immersed boundary method  proposed in~\cite{taira2007}, where  
the boundary forces $\boldsymbol{F}$ are treated as a Lagrange multiplier to ensure that the no-slip conditions $\boldsymbol{\xi} = \boldsymbol{\xi}_c$ at the cilia are satisfied. 
This method solves for the boundary forces  $\boldsymbol{F}$ implicitly with no need for additional constitutive relationship between $\boldsymbol{\xi}$ and $\boldsymbol{F}$. 
 The one-way coupled immersed boundary method is comparable in temporal stability to the classical fractional step method, thus enabling simulations with larger time steps \cite{taira2007}.

\begin{figure}[t]
	\centerline{\includegraphics{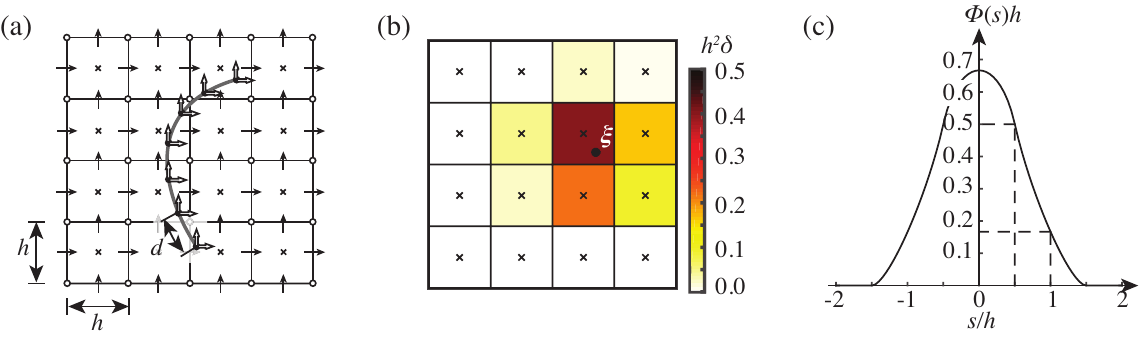}}
	\caption{{Numerical Methods. }(a) Staggered Eulerian mesh: the horizontal and vertical arrows ($\rightarrow,\uparrow$) represent the discrete velocities in $x,y$ directions respectively. The pressure $p$ and the normal stresses $\boldsymbol\sigma_{11},\boldsymbol\sigma_{22}$ are located at the cell centers ($\times$). The shear stress $\boldsymbol\sigma_{12}$ are located at the cell corners ($\circ$). The Lagrangian boundary points are represented by filled circles ($\bullet$). The discrete forces along the boundary are represented by thick arrows ($\Rightarrow,\Uparrow$).
	(b) The regularized Dirac delta function $\delta(\boldsymbol{x}(x,y)-\boldsymbol\xi(\xi,\eta)) = \phi(x-\xi)\phi(y-\eta)$ evaluated at the Eulerian mesh, where $\phi$ is a one-dimensional regularized Dirac delta function.
	(c) Here, we use the function $\phi$ employed by Roma \textit{et al.}~\cite{roma1999}.}
	\label{fig:mesh}
\end{figure}

We  embed a viscoelastic solver for the Oldroyd-B fluid Eqs.~(\ref{eq:govern},\ref{eq:govern2}) in the 
 one-way coupled immersed boundary method. To this end, we discretize the doubly-periodic fluid domain $\mathcal{F} = [0,L_x]\times[0,L_y]$ using a uniform, finite-volume, staggered Eulerian mesh, of mesh size $h$.
Details of the staggered mesh are shown in Fig.~\ref{fig:mesh}(a). 
At each time step, we explicitly update the elastic stress tensors $\boldsymbol{\sigma}_e$ in~\eqref{eq:govern2} using a standard second-order Runge-Kutta scheme.  Then, we substitute the  updated stress tensors in the momentum equation~\eqref{eq:govern} to update the flow field $\boldsymbol{u}$, subject to the incompressibility condition $\nabla \cdot \boldsymbol{u} = 0$ and no-slip boundary conditions~\eqref{eq:bc} using a fractional step method. 
In~\eqref{eq:govern}, we discretize the  convective term $\boldsymbol{u} \cdot \nabla \boldsymbol{u}$  using the explicit second-order Adams-Bashforth scheme and the diffusion term $\nabla^2 \boldsymbol{u}$ using the implicit Crank-Nicolson scheme.  The temporal and spatial discretization of equations~\eqref{eq:bc} and~\eqref{eq:govern} yields  an algebraic system of equations  for the fluid velocity $\boldsymbol{u}$. 

The cilium base point is located at $(x,y) = (L_x/2,0)$ and the effective stroke is pointing in the positive $x$-direction as shown in Fig.~\ref{fig:compmodel}(b).  We use the mesh size $d=h$ to discretize the cilium, as recommended by \cite{taira2007} to ensure no penetration of streamlines.
To communicate between the fluid domain and the moving cilia boundary, i.e., between the Eulerian and Lagrangian meshes, we discretize the Dirac delta function in~\eqref{eq:interp1} and~\eqref{eq:interp2} using the discrete Dirac delta function developed by Roma \textit{et al.}~\cite{roma1999} and illustrated in Figs.~\ref{fig:mesh}(b) and (c). 
 The forces $\boldsymbol{F}$  and  fluid velocity $\boldsymbol{u}$ are updated simultaneously  at each time step. 

\begin{figure}[!t]
	\centerline{\includegraphics{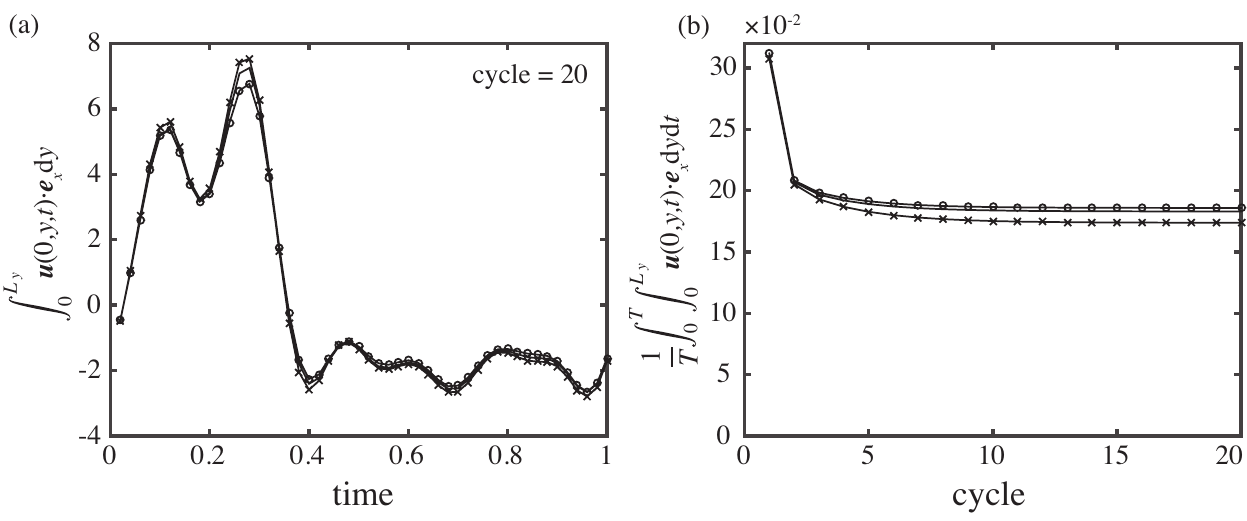}}
	\caption{{Numerical Convergence.} ({a}) The flow rate within the 20th cycle. ({b}) The mean flow rate  as a function of cycles. The Deborah number of the mucus layer is $\text{De}_m=5$, the periciliary layer thickness is $L_p=0.8$. `$-$': $h = 0.0156$, $\Delta t = 2\times10^{-5}$; `\sout{$\times$}': $h = 0.0312$, $\Delta t = 2\times10^{-5}$; `\sout{$\circ$}': $h = 0.0156$, $\Delta t = 4\times10^{-5}$.  
	}
	\label{fig:conv} 
\end{figure}

We let the fluid domain $\mathcal{F}$ be of size $L_x \times L_y = [0,2] \times [0,2]$, and we set the total integration time to be $20$ cilia-beating cycles.
The numerical values of the  mesh sizes $h = d = 0.0156$ and the timestep $\Delta t = 2\times 10^{-5}$ are chosen so that the solution changes little under further  spatial and temporal mesh refinement. In particular, a decrease in the spatial and temporal  mesh size by a factor of 2 yields changes in the terminal cilia-driven flow rate of about $5\%$ (under spatial mesh refinement) and $1\%$ (under temporal mesh refinement), see Fig.~\ref{fig:conv}. Therefore, we consider the discretization scheme to have numerically converged.

 In all our simulations, we set the Reynolds number to be $\text{Re} = 0.1$, the elastic viscosity fraction $\alpha = 0.5$, and the Oldroyd-B regularization parameter $\epsilon = 0.05$. We consider a small Deborah number De$_p= 0.05$ of the periciliary layer (nearly Newtonian) and we vary the elastic properties De$_m$ of the mucus layer and its thickness $L_m$, which implies a change in the thickness of the periciliary layer $L_p$.


\section{Results}
\label{sec:results}

We consider two states of periciliary layer thickness: $L_p = 0.8$, which we label as a ``healthy" state because it is comparable to the average length of the cilium over its beating cycle as observed experimentally in healthy ciliary systems, and $L_p = 0.4$, which we label as a ``diseased" state. In the healthy state, the cilium tip penetrates into the mucus layer during the effective stroke as it pushes the mucus forward while the whole cilium moves in the periciliary layer during the recovery stroke. In the diseased state, the mucus layer covers part of the cilium at all times, even during the recovery stroke.

\begin{figure}[!t]
	\centerline{\includegraphics{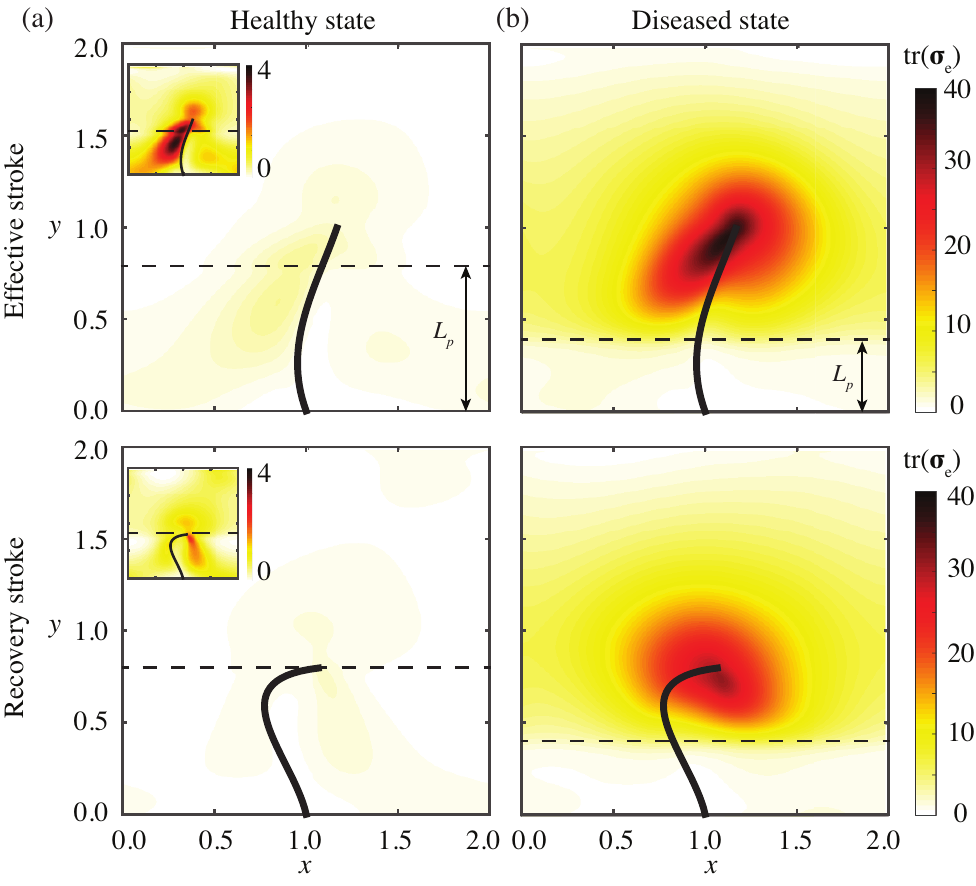}}
	\caption{{Elastic stress energy.} Contour plots of the stress energy $\text{tr}(\boldsymbol{\sigma}_e)$ in  (a) healthy ($L_p = 0.8$) and (b) diseased ($L_p = 0.4$) states, shown at $t = 0.2T$ during the effective stroke and $t= 0.7T$ during the recovery stroke. Here, $\text{De}_m=5$  and De$_p = 0.05$. 
	}
	\label{fig:stress} 
\end{figure}

Fig.~\ref{fig:stress} shows the contour plots of the elastic stress energy in these two states for mucus Deborah number De$_m=5$. The elastic stress energy is  equal to the trace of the extra stress tensor tr$(\boldsymbol{\sigma}_e)$. Because the mucus layer has a longer relaxation time than the periciliary layer, the cilium in the diseased state generates higher fluid stresses (about 10 times) compared to the cilium in the healthy state.
The spatial distribution of the elastic stress energy is also different in the two states.  During the effective stroke, the stress is distributed along the cilium in the healthy state, while in the diseased state it is concentrated close to the  cilium tip. During the recovery stroke, we observe larger regions of high stress with slower decay rate in the diseased state  compared to the healthy state. We next examine the effects of the elastic stress in the mucus on the fluid velocity field in both the health and diseased environments.

\begin{figure*}[t]
	\centerline{\includegraphics{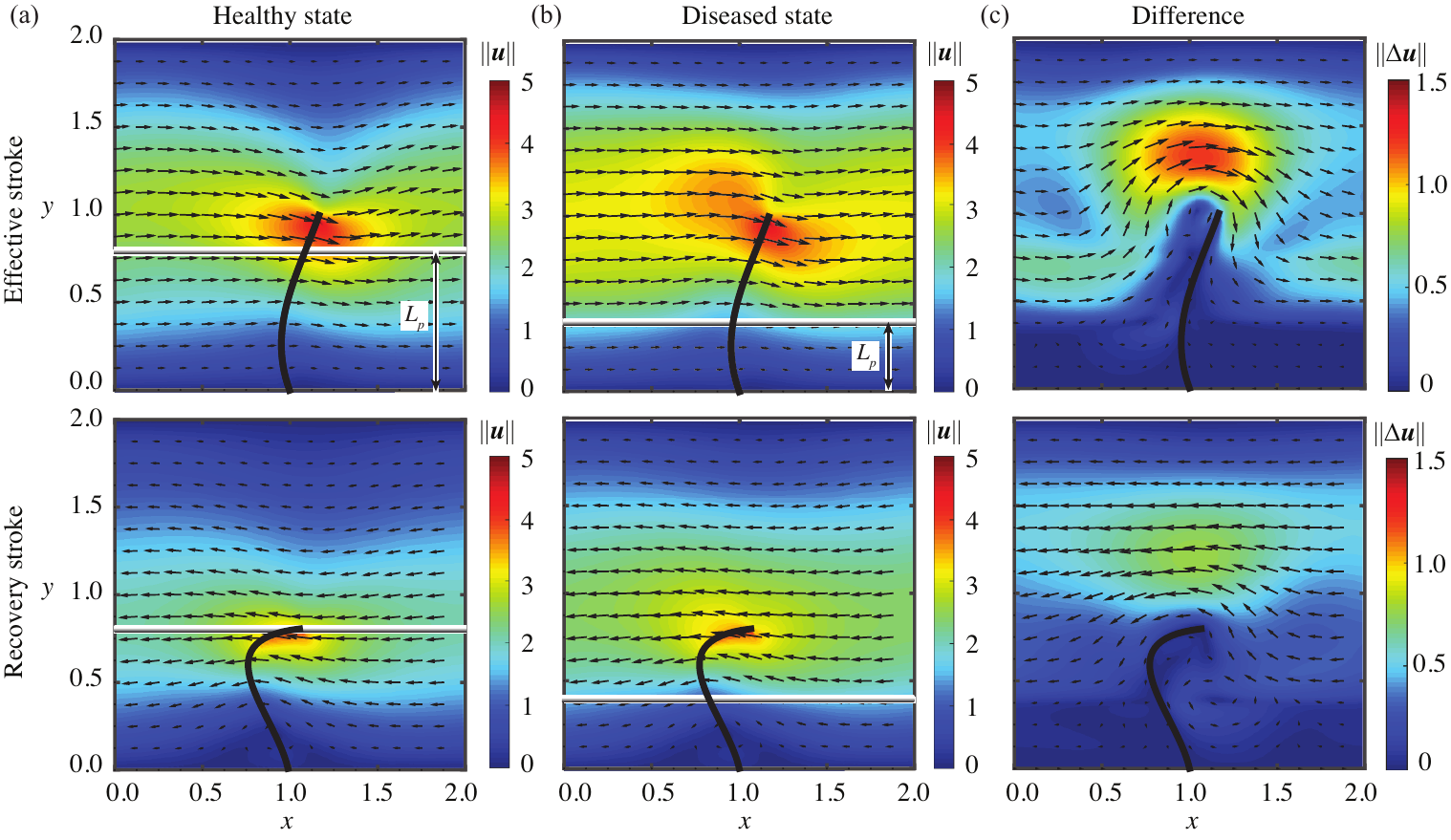}}
	\caption{{Velocity vector fields.} Velocity fields  in ({a}) healthy state ($L_p = 0.8$) and ({b})  diseased state ($L_p = 0.4$). ({c}) The difference between the two vector fields. The colormap shows  the magnitude of the fluid velocities. Here, $\text{De}_m=5$,  De$_p = 0.05$. 
	}
	\label{fig:vel} 
\end{figure*}

\subsection{Cilia-driven Fluid Velocity}

Figs.~\ref{fig:vel}({a}) and ({b}) depict  the flow fields of the cilium  in the healthy and diseased states respectively. The fluid velocities close to the cilium are dictated by the cilium motion because of the no-slip boundary conditions. However, the decay of  the flow velocities over space  is affected by the elastic stress distribution. Compared to the healthy state, the velocities decay slower in the diseased state because of the high level of stress energy. The velocity differences between the two states $\Delta \boldsymbol{u} = \boldsymbol{u}_{\text{diseased}}-\boldsymbol{u}_\text{healthy}$ are shown in Fig.~\ref{fig:vel}({c}).  It shows an  increase in positive flow during the effective stroke and an increase in negative flow during the recovery stroke, with the maximum difference in velocity appearing above the tip of the cilium. 
This instantaneous picture is not sufficient to assess the overall velocity difference between the healthy and diseased states. For that, we evaluate the time averaged flow fields over one beating cycle, as discussed next.

\begin{figure*}[t]
	\centerline{\includegraphics{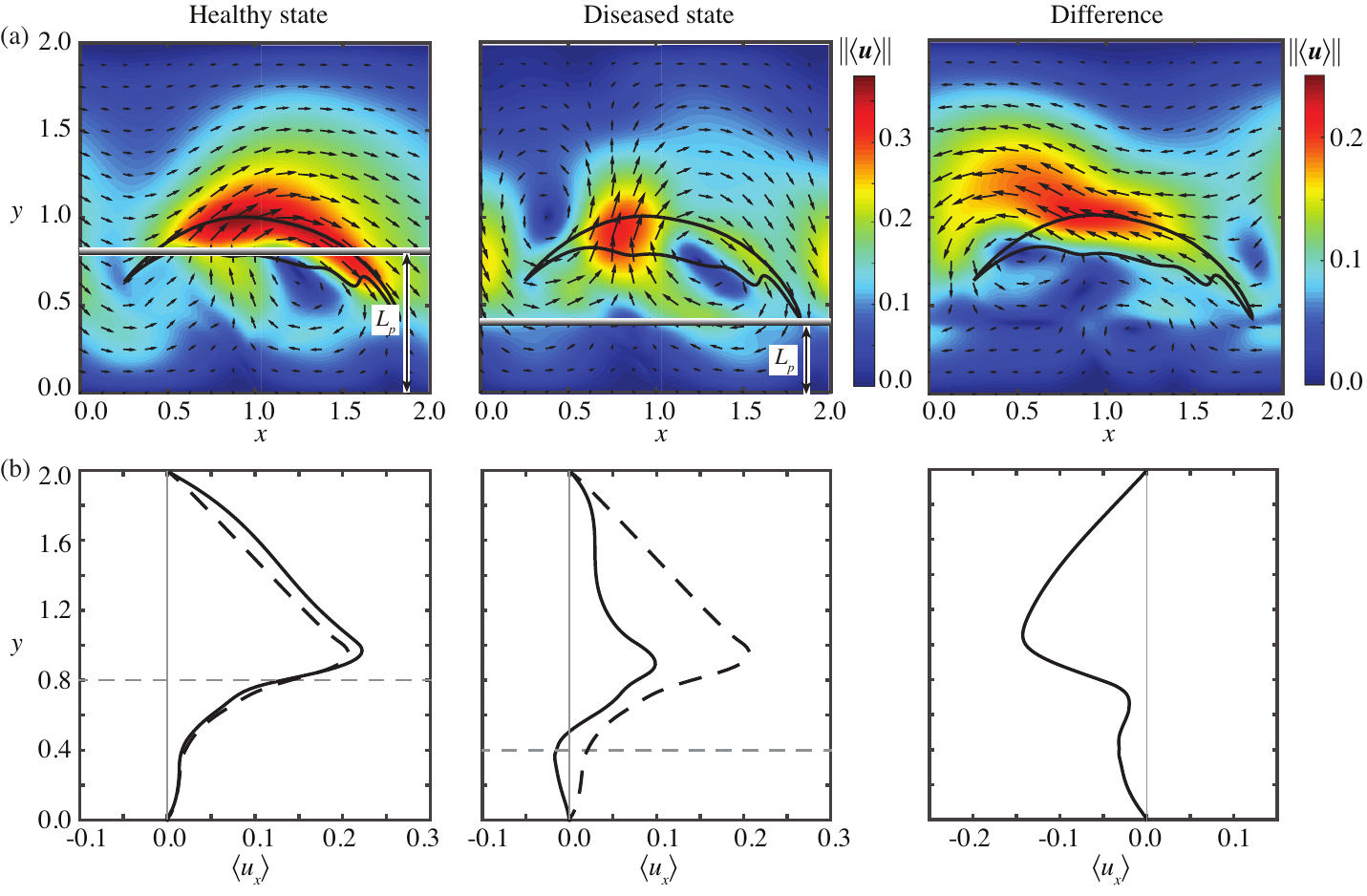}}
	\caption{{Flow velocity averaged over one cilia beating cycle.}  ({a}) Mean velocity fields over one beating cycle. The closed loops represent the trajectories of the cilium tip over one cycle. ({b}) Mean flow profiles in $x$-direction as functions of $y$. The black dashed curves represent the flow profile of a control state where there is no mucus ($L_p = 2.0$). The light grey dashed lines denote the interface between the mucus/periciliary layers. }
	\label{fig:velmean} 
\end{figure*}

Fig.~\ref{fig:velmean}({a}) shows the time averaged flow fields $\langle \boldsymbol u \rangle = \frac{1}{T}\int_0^T\boldsymbol u \mathrm{d}t$. In the healthy state, the velocity field averaged over one cycle shows a single vortex-like structure below the cilium tip.  On the other hand,  $\langle \boldsymbol u \rangle$ in the diseased state shows two counter-rotating vortex-like structures. Particularly, the vortex structure at the front end (left) of the cilium rotates counter-clockwise, generating a back flow above the cilium tip in this location. It is peculiar that the healthy state favors an asymmetric flow structure while the diseased state is characterized by a symmetric flow. Symmetry here is detrimental to net flow and transport in the direction of the effective stroke.

We further average the velocity field in the $x$-direction, namely, we compute $\langle u_x \rangle = \frac{1}{L_x}\int_0^{L_x} \langle \boldsymbol u \rangle \cdot \boldsymbol{e}_x \mathrm{d}x$, with $\boldsymbol{e}_x$ being the unit vector pointing to the $x$-direction. The average velocity profiles are depicted  in Fig.~\ref{fig:velmean}({b}) and correspond to nonlinear shear profiles.  
The results of a benchmark computation in which the periciliary layer takes up the entire channel (no mucus) are shown in dashed lines. We will refer to this benchmark computation as the {\em control case}. In the healthy state, the velocity $\langle u_x \rangle$ is almost identical to that of the control case for $y<0.8$ (periciliary layer) and  faster than that of the control state  for $y>0.8$ (mucus layer), with $\max(\langle u_x \rangle)_{\rm healthy} = 0.22$. In the diseased state, the flow velocity lags behind the control case for all $y$ and, unlike the healthy and control states, the flow is negative in the periciliary layer. The highest mean velocity   in the diseases state is $\max(\langle u_x \rangle)_{\rm diseased} = 0.10$, that is, $\max(\langle u_x \rangle)$ drops by over $50\%$ from the healthy.
This drop in fluid transport can be attributed to the higher stress energy in the diseased state, which makes the mucus layer ``stiffer" and solid like, thus hindering fluid velocity and flow transport.

\begin{figure}[t]
	\centerline{\includegraphics{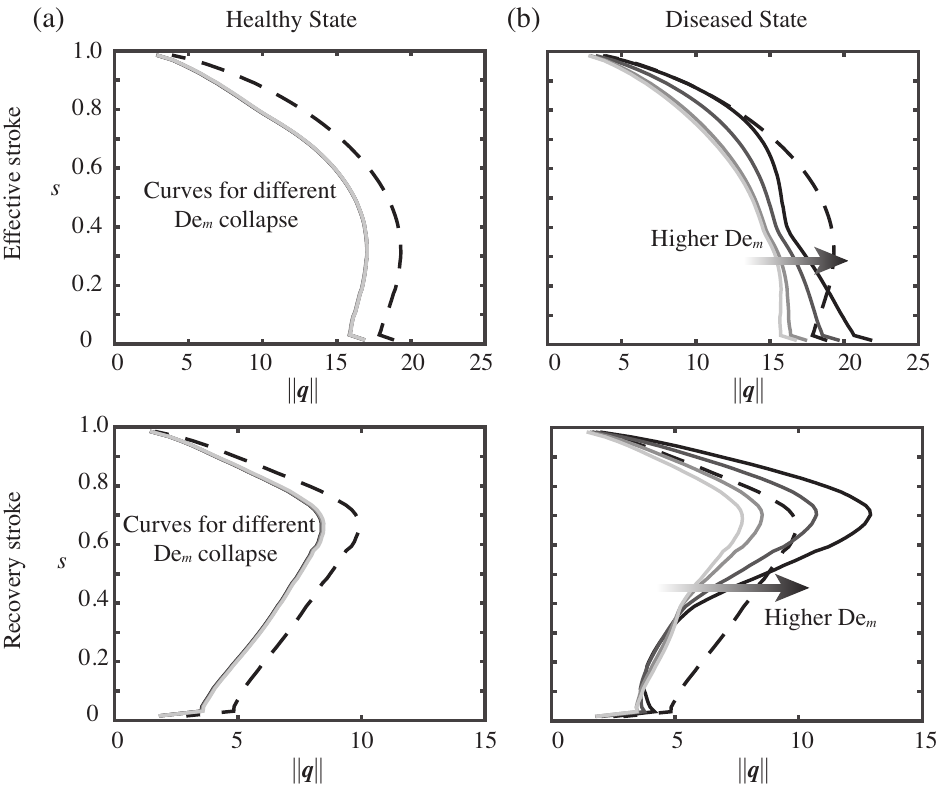}}
	\caption{{Internal cilia torques.} Internal torques in (a) healthy ($L_p = 0.8$) and  (b) diseased ($L_p = 0.4$) states 
	as functions of the arclength along the cilium for Deborah numbers $\text{De}_m = 1,2,5,10$. Snapshots during the effective  	($t=0.2$) and recovery ($t = 0.7$) strokes. 
	 The dashed curves represent the results of the control case with no mucus ($L_p = 2.0$).}
	\label{fig:torques5} 
\end{figure}

\subsection{Cilia Internal Moments and Power Requirements}
We now evaluate the  internal moments required in order for the cilia to perform the prescribed beating kinematics in both the healthy and diseased conditions. We assume that the beating kinematics is not affected by the thickness of the mucus layer or its elastic properties.  This assumption is useful to compare the  performance of the unchanged cilia kinematics under various mucus conditions. The internal bending moments $\boldsymbol{q}$ generated by each cilium are obtained using the Kirchhoff model for an elastic filament as done in~\cite{eloy2012, guo2014}, which yields
$\boldsymbol{q} = B\boldsymbol{t}''  \times\boldsymbol{t}+\boldsymbol{t}\times\int_s^l\boldsymbol{F}(\tilde{s},t)\mathrm{d}\tilde{s}.$
Here, $B$ is the dimensionless bending rigidity of the cilium, which we set to $B =  6.54\times 10^{-4}$ as done in \cite{eloy2012,guo2014}, $\boldsymbol{t}$ is the unit tangent to the cilium and the prime superscript $(\cdot)'$ represents derivative with respect to arclength $s$ ($\boldsymbol{t}'' = {\partial^2 \boldsymbol{t}}/{\partial s^2}$).

Fig.~\ref{fig:torques5} depicts the magnitude of the internal bending moments $\| \boldsymbol{q} \|$ as a function of the arclength $s$ at two snapshots corresponding to $t = 0.2T$  during the effective stroke and $t = 0.7T$ during the recovery stroke. Fig.~\ref{fig:torques5}({a}) shows the values of $\| \boldsymbol{q} \|$ in the healthy state ($L_p = 0.8$) for four different values of De$_m= 1, 2, 5, 10$ compared to the control case (no mucus, dashed line) and Fig.~\ref{fig:torques5}({b}) compares the diseased state ($L_p = 0.4$, De$_m= 1, 2, 5, 10$) to the control case. Interestingly, in the healthy state, the magnitude $\| \boldsymbol{q} \|$ of the internal moments  is independent of De$_m$ and it takes smaller values than those in the absence of mucus (dashed line). That is to say, in the healthy state and within the considered range of Deborah numbers, the internal moments required to perform the cilia beating kinematics are independent of the elastic properties of the mucus; and the same cilia beating kinematics require weaker internal moments in the presence of mucus than in its absence. The latter can be explained as a result of the interplay between the elastic properties of the fluid and the oscillatory motion of the cilia. Elasticity causes the fluid to ``react" in the opposite direction once the applied forces are released.  Therefore, at each reversal in the cilium motion from effective to recovery stroke and vice versa, the elastic reaction of the fluid tends to reinforce the reversal in the cilium motion, thus requiring the cilium to exert smaller forces on the fluid and resulting in smaller internal moments along the cilium.
On the other hand, in the diseased state, the required internal moments  increase as a function of the mucus Deborah number and exceed the internal moments of the control case at $\text{De}_m = 5$ and $10$. In the diseased state, as the Deborah number increases, the mucus becomes more stiff with little  difference in the elastic energy between the effective and recovery strokes (Fig.~\ref{fig:stress}). This causes the system to loose the favorable interplay observed in the healthy state between fluid elasticity and oscillatory cilia kinematics and requires higher internal moments for the cilium to complete its beating cycle.

The average power $\langle P \rangle$ expended internally by the cilium is equal to the power consumed by the internal moments $\boldsymbol{q}$. Namely, one has
$\langle P \rangle = \langle \int_0^L \max(0,\boldsymbol{q}\cdot\boldsymbol{\Omega})\text{d}s\rangle,$
where 
$
\boldsymbol\Omega = \left( \| \boldsymbol{\dot{t}}\| /{\| \boldsymbol{t}\times\boldsymbol{\dot{t}}\|}\right) {\boldsymbol{t}\times\boldsymbol{\dot{t}}}
$ 
is the angular velocity vector and the dot represents derivative with respect to time ($\dot{\boldsymbol{t}} = {\partial \boldsymbol{t}}/{\partial t}$). Negative work is not taken into account because the cilium does not harvest energy from the ambient flow~\cite{eloy2012}.
\textcolor{black}{Since the internal moment required in the healthy state is lower than that in the control case (no mucus), the resulting power is lower as well ($\langle P \rangle = 72.5$ in the healthy state vs. $81$ in the diseased state).} This reduction in power requirement is consistent with the analysis of Lauga~\cite{lauga2007}, where he  examined analytically  the swimming motion of  a small amplitude waving sheet in weakly-viscoelastic fluids. Lauga showed that it is energetically beneficial to swim in a fluid that has some elasticity, compared to a Newtonian fluid with the same viscosity. The power requirement in the diseased state with $\text{De}_m > 8$ is higher than that of the control case. In this case, Lauga's assumption of weak viscoelasticity is no longer valid.

\begin{figure*}[t]
	\centerline{\includegraphics{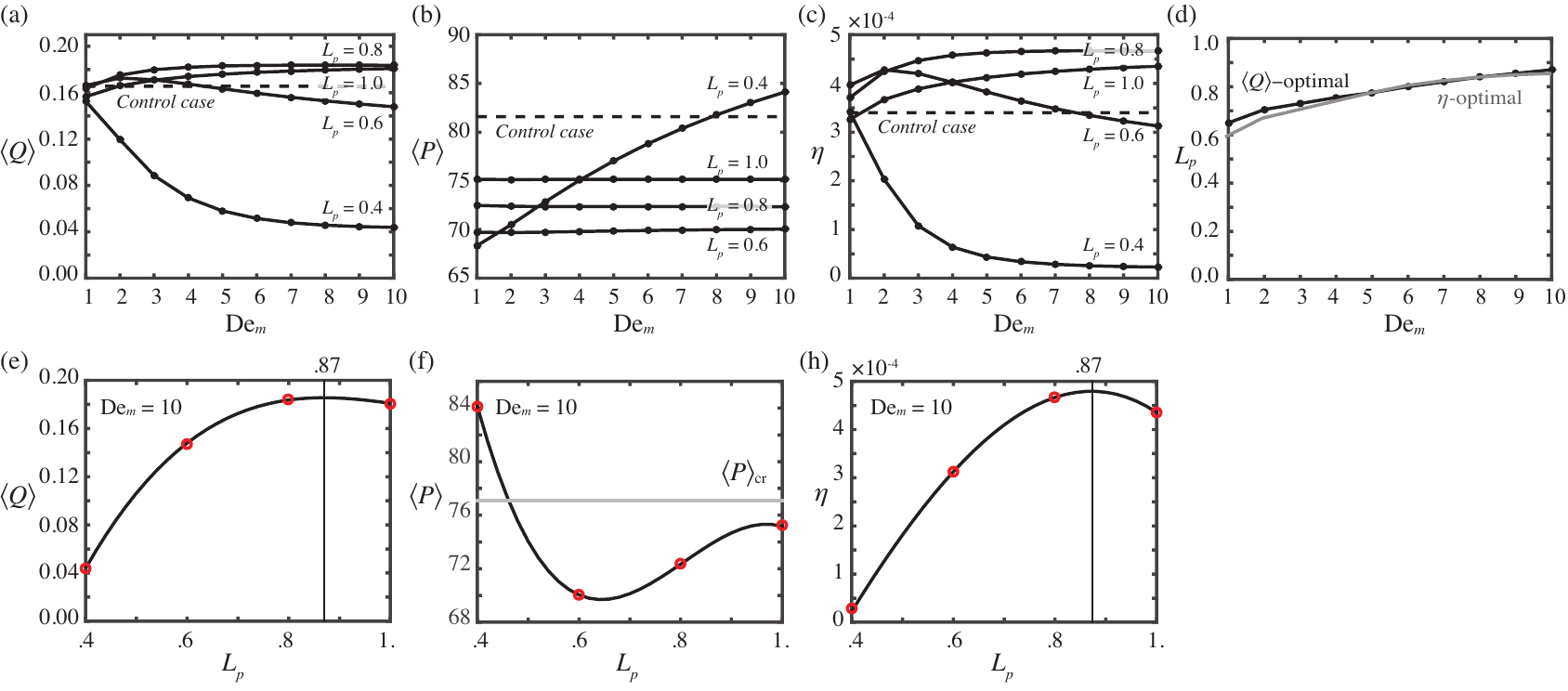}}
	\caption{{Cilia performance in terms of the periciliary layer thickness $L_p$ and mucus relaxation time $\text{De}_m$.} 
	(a) mean flow rates, (b) mean internal powers, and (c) transport efficiencies for different periciliary layer thicknesses as functions of the relaxation time of the mucus layer. Horizontal dashed lines show the results of the control case of no mucus ($L_p = 2.0$). (d)
	 $\langle Q\rangle$- and $\eta$-optimal values of $L_p$  as a function of $\text{De}_m$.
	(e) mean flow rates, (f) mean internal powers, and (h) transport efficiencies as functions of periciliary layer thickness $L_p$  for $\text{De}_m=10$. Results are interpolated by cubic spline functions. Grey vertical lines show the periciliary layer thicknesses with the highest flow rate and transport efficiency. $\langle P \rangle_{\text{cr}}$ represents the (presumed) energy budget of cilia. } 
	\label{fig:results} 
\end{figure*}

\subsection{Effects of Mucus  Layer  on Flow Transport, Power and Efficiency}
We study the effects of the Deborah number De$_m$ and thickness $L_m$ of the mucus layer on the mean flow transport $\langle Q\rangle$, mean power expenditure $\langle P\rangle$ and efficiency $\eta$ of the cilia. 
The total flow transported by the cilia in the channel is equal to the flow across any plane at constant $x$ (a direct result of flow incompressibility). That is to say, the flow rate at any given time is given by $Q(t) = \int_0^{L_y} \boldsymbol{u}(0,y,t)\cdot\boldsymbol{e}_x\mathrm{d}y$. Averaging over one cycle, we get the mean flow rate $\langle Q \rangle = \frac{1}{T}\int_0^TQ\mathrm{d}t$. 
Given $\langle Q\rangle$ and  $\langle P \rangle$, we define  the cilia transport efficiency as 
$\eta = \mu l^{-3}{\langle Q\rangle^2}/{\langle P \rangle}$.
This definition of efficiency is consistent with that employed in \cite{eloy2012,osterman2011}.

We vary the Deborah number of the mucus layer
De$_m$ from $1$ to $10$. 
Fig.~\ref{fig:results}({a}) shows the rate of flow transport  as a function of the Deborah number for four distinct  values of the periciliary layer thickness:
$L_p = 0.4, 0.6, 0.8$ and $1$. The values of $L_p = 0.4$ and $0.6$ correspond to diseased conditions where the mucus layer penetrates into the periciliary space.
Evidently, the effects of the mucus elasticity depends highly on the thickness of the mucus layer. Higher elasticity increases the flow rate in the healthy states while decreases the flow rate in the diseased states.  The flow rates of the healthy states are higher than the control case (no mucus, dashed line) while those of the diseased states are lower. In fact, the flow rate of the diseased state $L_p = 0.4$ is only $20\%$ of the  healthy flow rates at $\text{De}_m = 10$.
These findings are consistent with the results of Teran \textit{et al.}~\cite{teran2010}, where they showed that compared to the viscous case, elasticity can be beneficial to the swimming of a sperm cell for a range of Deborah number and detrimental to the swimming motion if the Deborah number is too high. Further, Fig.~\ref{fig:results}(a) also shows that for a  given  thickness of the periciliary layer, the flow rates plateau for $\text{De}_m~\approx 10$, suggesting that the transport performance is robust to further changes in viscoelastic properties of the mucus layer. 

The mean power $\langle P \rangle$ expended by the cilium in one cycle is depicted in Fig.~\ref{fig:results}({b}). Clearly, for $L_p\ge 0.6$ the power expenditure is almost independent of the mucus Deborah number. However, for  $L_p = 0.4$, the power requirement increases with De$_m$. That is to say,  the thinner periciliary layer $L_p = 0.4$ is characterized not only by a lower flow rate $\langle Q \rangle$ but also by a higher power requirement. Therefore, when limited by the energy budget of cilia in real biological system $\langle P\rangle_{\text{cr}}$, cilia may not be able to complete the beating cycle once the required power exceeds that budget.

Without access to the energy budget of biological cilia, we assume that the cilia have enough power to complete the prescribed beating cycle and compute the transport efficiencies (see Fig.~\ref{fig:results}({c})). The curves exhibit similar trends as the flow rates $\langle Q \rangle$. In the healthy states, higher efficiencies are observed for higher Deborah numbers. In the diseased states, the efficiencies decrease  with increasing Deborah numbers. In particular, at $\text{De}_m = 10$, the efficiency of $L_p = 0.4$ is only $5\%$ of the efficiency of $L_p = 0.8$. 
\textcolor{black}{We further plot the values of $L_{p}$ that maximize $\langle Q\rangle$ and $\eta$ versus the mucus Deborah number $\text{De}_m$ in Fig.~\ref{fig:results}(d). Clearly, the optimal $L_{p}$ increases from $0.6$ to $0.9$ as $\text{De}_m$ increases from $1$ to $10$ but at a decreasing rate of change for larger $\text{De}_m$. That is to say, for larger $\text{De}_m$, the optimal periciliary layer thickness seems to level off and is not very sensitive to the further changes in $\text{De}_m$.} 

Finally, we highlight the dependence of the performance metrics $\langle Q \rangle$, $\langle P \rangle$ and $\eta$ on the thickness $L_p$ of the periciliary layer  for a fixed value of De$_m = 10$. The flow rate $\langle Q \rangle$ is not a monotonic function of  $L_p$ (Fig.~\ref{fig:results}(e)). It is largest for $L_p = 0.87$, for which the efficiency $\eta$ is also largest (Fig.~\ref{fig:results}(h)). That is, $L_p = 0.87$ is the optimal value of the periciliary layer thickness. This value is close to the value $L_p = 0.8$, which we labeled as healthy state, where the thickness of the periciliary layer is comparable to the average cilia length. 
On the other hand, the required power reduces gradually when $L_p$ decreases from $1$ to $0.6$, but increases sharply when $L_p <0.6$ (see Fig.~\ref{fig:results}({f})). 
 An increase in the mucus layer $L_m$ and a decrease in the periciliary thickness  $L_p$ may require higher powers than those affordable by the cilia, thus preventing the cilia from completing their beating cycle, leading to decreased transport performance and even complete failure in their transport function, as observed in lung diseases including CF and COPD. 
 


\section{Discussion}
\label{sec:discussion}
Mucociliary clearance in the lung serves to effectively transport inhaled toxic molecules and undesirable particles away from the  tissue surfaces, thus shielding the airways from potentially infectious agents. 
Disruptions in the ciliary apparatus, whether due to a genetic disorder
or acquired causes, are directly linked to infection and disease such as CF and COPD.
In these diseased conditions, the mucociliary system is characterized by a depletion of the watery periciliary layer underlying the mucus layer.
Clinical evidence connects the periciliary layer depletion to reduced rates of mucus clearance~\cite{fahy2010,boucher2007}. \textcolor{black}{In fact, \textit{in vivo} analyses of wild-type (healthy) and CF murine nasal airway surface liquid  thickness show that CF decreases the airway surface liquid height from $7\mu \text{m}$ to $4\mu \text{m}$, with periciliary layer thickness decreased almost by half from $4\mu \text{m}$ to $2.5\mu \text{m}$~\cite{tarran2001}.}
 
In this work, we developed a novel computational model to study mucociliary transport in a microfluidic channel consisting of a mucus layer (viscoelastic fluid) atop a periciliary layer (nearly-viscous fluid). 
We systematically varied the viscoelastic properties and thickness of the mucus layer to emulate healthy and diseased conditions. \textcolor{black}{The representative diseased state has a periciliary layer thickness half of the representative healthy state, which is consistent with experimental observations~\cite{tarran2001}.} We assessed cilia performance in terms of  three metrics:  flow transport, internal power expended by the cilia, and transport efficiency. We found that, compared to a control case with no mucus, a healthy mucus layer enhances cilia performance in all three metrics. That is to say, a  layer of mucus atop a healthy periciliary layer not only improves flow transport but it does so at an energetic advantage for the cilia.  Further, in healthy environments,  increasing the Deborah number of the mucus layer enhances transport efficiency. 
In contrast, in diseased environments where the periciliary layer is depleted, mucus hinders transport and larger Deborah numbers reduce transport  efficiency further.  This decrease in efficiency is accompanied by an increase in the internal torques and power  needed  to complete the cilia beating cycle. 
Cilia therefore may not be able to beat at all if the required power is higher than the power afforded by the cilia internal machinery - this is consistent with clinical observations that link thin periciliary layers  to cilia failure and dysfunction \cite{fahy2010}. It is worth noting here that 
the transport efficiency we obtained in the diseased state of a depleted periciliary layer $L_p = 0.4$ 
is only $5\%$ of the value of the transport efficiency obtained in the healthy state of a periciliary layer thickness $L_p = 0.8$. This decrease in performance in the diseased state is accompanied by larger elastic stresses at the cilia tips and by  two counter-rotating vortex-like structures below the cilia tips.  The symmetry of these dipolar vortex structures is detrimental to net transport. 

Our modeling framework systematically couples mucociliary transport to the parameters of the mucus layer. It establishes  variations in the flow transport in response to perturbations in the mucus and periciliary layers, regardless of the physiological mechanisms that bring this about. Here, a few comments on the model advantages and limitations are in order:

\setlist[itemize,1]{leftmargin=0cm}
\begin{itemize} \itemsep -0.5\baselineskip 
\item[] (i) Our underlying assumption throughout this work was that the beating kinematics is not affected by the thickness of the mucus layer or its elastic properties.  This framework is useful to compare the  performance of the unchanged cilia kinematics under various mucus conditions. We found that cilia performance is enhanced in the presence of a healthy mucus layer than in the control case of no mucus, and diminished in diseased environments when the periciliary layer is depleted. A future extension of the present study would be to allow the cilia beat kinematics to change dynamically in response to changes in the mucus environment. Such complementary approach would provide valuable insights into the effects of diseased environments on changes in the cilia beating patterns. 

 \item[] (ii)  All cilia in this study are beating in synchrony, as ensured by the periodic boundary conditions in the $x$-direction. In reality, cilia beat in an orchestrated metachronal wave. In a previous work that does not explicitly account for the mucus layer, we showed that compared to beating in synchrony, a metachronal beat can result in as much as 6 times higher flow velocity for this particular cilia beating pattern~\cite{guo2014}. Taking this as a guideline, we scale the dimensionless average flow rate in the healthy condition $\langle Q \rangle / L_y = 0.09$ by the characteristic length $l_c = 6\times 10^{-6}$ m and time $t_c = 1/15$ s reported in~\cite{fulford1986} for the cilia beating pattern we considered in this study. The resulting estimate of the flow rate under healthy conditions is about $48.6~\mu$m/s in our model. 
 Our estimate is congruent with the estimate obtained in~\cite{smith2007viscoelastic} and shows good agreement to various experimental measurements. In particular, experimental flow rates in the range  $70 - 92~\mu$m/s were reported in~\cite{icrp1994}  for tracheal transport of healthy subjects and in the range  $67 - 333\mu$m/s in \cite{salathe1997mucociliary} using less invasive measurement technique which yield lower flow rates. More recently, a flow rate of $39.2\mu$m/s was reported in~\cite{matsui1998coordinated}.

\item[] (iii) Mucus layers are characterized by large Deborah numbers, of order $10-100$ in healthy environments and even larger in diseased states{~\cite{lauga2007, gilboa1976}}. Here, we varied the Deborah number De$_m$ from 1 to 10 to establish a trend of how performance depends on De$_m$. We found that, in healthy environments, the transport rates increase with increasing De$_m$ and reach a plateau as De$_m$ approaches 10, whereas under diseased conditions, the transport rates decrease with increased  De$_m$ before reaching their plateau value ({Fig.~\ref{fig:results}}). We also found that the internal moments along the cilia are not very sensitive to De$_m$ under healthy conditions, but sharply increase with  De$_m$ when the periciliary layer is depleted. These findings justify the choice of the range of De$_m$ considered in this study. In the future,  larger De$_m$ will be considered, which will render the system of equations ``stiff"  and more challenging to solve computationally, requiring the use of a fully implicit fluid solver. 

\item[] (iv) Our results are based on two-dimensional (2D) computations and 2D
cilia beating kinematics. We used this 2D set-up to
better illustrate the main ideas and for easier visualization of the resulting 
flows. In this set-up, a cilium should be thought of as a ``wall" of synchronously beating cilia in the third $z$-direction (perpendicular to the $(x,y)$-plane). It is worth noting that the computational framework we used is general and can be easily extended to include 3D cilia beating patterns, albeit at an increased computational cost. 
\end{itemize}

The two main results obtained from this study --  the fact that a healthy mucus layer enhances the performance of mucociliary systems and that a depleted periciliary layer can be directly linked to diminished transport and cilia dysfunction through excessive demands on cilia internal moments --  serve to complement ongoing research on understanding cilia-related diseases and to direct future studies. In conjunction with its role in understanding cilia-related diseases, the quantitative model we presented in this work could play important roles in the design and use of microfluidic devices in health-related cilia research. Indeed, a novel and exciting research direction in {\em in vitro} cell cultures lies in the development of engineered ciliated tissues in microfluidic chips, so called `organs-on-chips', as  the next-generation platforms for basic research, drug development, and diagnostics{~\cite{benam2015}}. Traditionally,  the clinical use of {\em in vitro} cell cultures in health-related research on respiratory tissues and mucociliary transport~{\cite{seybold_mucociliary_1990, wanner_mucociliary_1996}} has been mostly qualitative and lacking a direct translation to clinical readouts.
Organs-on-chips aim to provide clinically relevant metrics of ciliated tissue health by recapitulating and quantifying essential structure-function relationships~\cite{nawroth2013},  thereby achieving better predictions of disease mechanisms and treatment options in humans compared to traditional cell culture and animal models~\cite{huh2010, bhatia2014microfluidic}. Our three quantitative measures of cilia transport performance can be applied to experimental data obtained from cilia-on-chip systems, thus opening the door to direct quantitative comparisons of {\em in-vitro} healthy and diseased cilia conditions. Quantitative models such as the one presented here will be important tools for understanding the link between tissue-engineered ciliated organs and functional outcomes. 

Another future direction of this work that is directly relevant to infectious disease of the airways is to determine what is required by a beneficial or pathogenic bacterial cell that colonizes the airway surfaces  to resist or to work within the  healthy environment of mucociliary clearance. Clearly to successfully colonize the ciliated surface, such cells must have developed mechanisms to avoid mucociliary clearance. Most research has focused on the chemical properties of the bacteria-tissue interactions, see, e.g.,~\cite{mcfallngai2014} and references therein, whereas several recent studies in biofluid mechanics have considered  bacterial motility in otherwise still viscoelastic environments~\cite{thomases2014, teran2010, lauga2007, qin2015, patterson2015}. However, the mechanical strategies employed by motile bacterial cells in complex environments such as in healthy and diseased mucociliary clearance remain an open research problem. 

Finally, we note that in addition to their importance in clinical applications, the efficiency of biological cilia in pumping and mixing fluids at very small scales provided an attractive paradigm for fluid manipulation by artificial cilia at the micron scale~\cite{evans2007magnetically,khatavkar2007active, den2008artificial, vilfan2010self}. 
Also, self-propelled microrobots by flagellar or ciliary activities are being proposed as revolutionary devices in the field of minimally invasive medicines~\cite{nelson2010microrobots, peyer2013bio}. The quantitative model we presented in this study for assessing cilia performance in complex environments could serve as an important design tool in such microfluidic applications.

\begin{acknowledgments}
The authors acknowledge partial support from NSF under an INSPIRE grant and are indebted to Prof. Margaret McFall-Ngai and Dr. Janna Nawroth for many useful discussions.
\end{acknowledgments}

\bibliography{references}

\end{document}